\documentclass[conference]{IEEEtran}
\IEEEoverridecommandlockouts
\usepackage{cite}
\usepackage{amsmath,amssymb,amsfonts}
\usepackage{algorithmic}
\usepackage{graphicx}
\usepackage{textcomp}
\usepackage{xcolor}
\usepackage{hyperref}
\newcommand{\orcidicon}[1]{%
    \href{https://orcid.org/#1}{%
        \includegraphics[width=10pt]{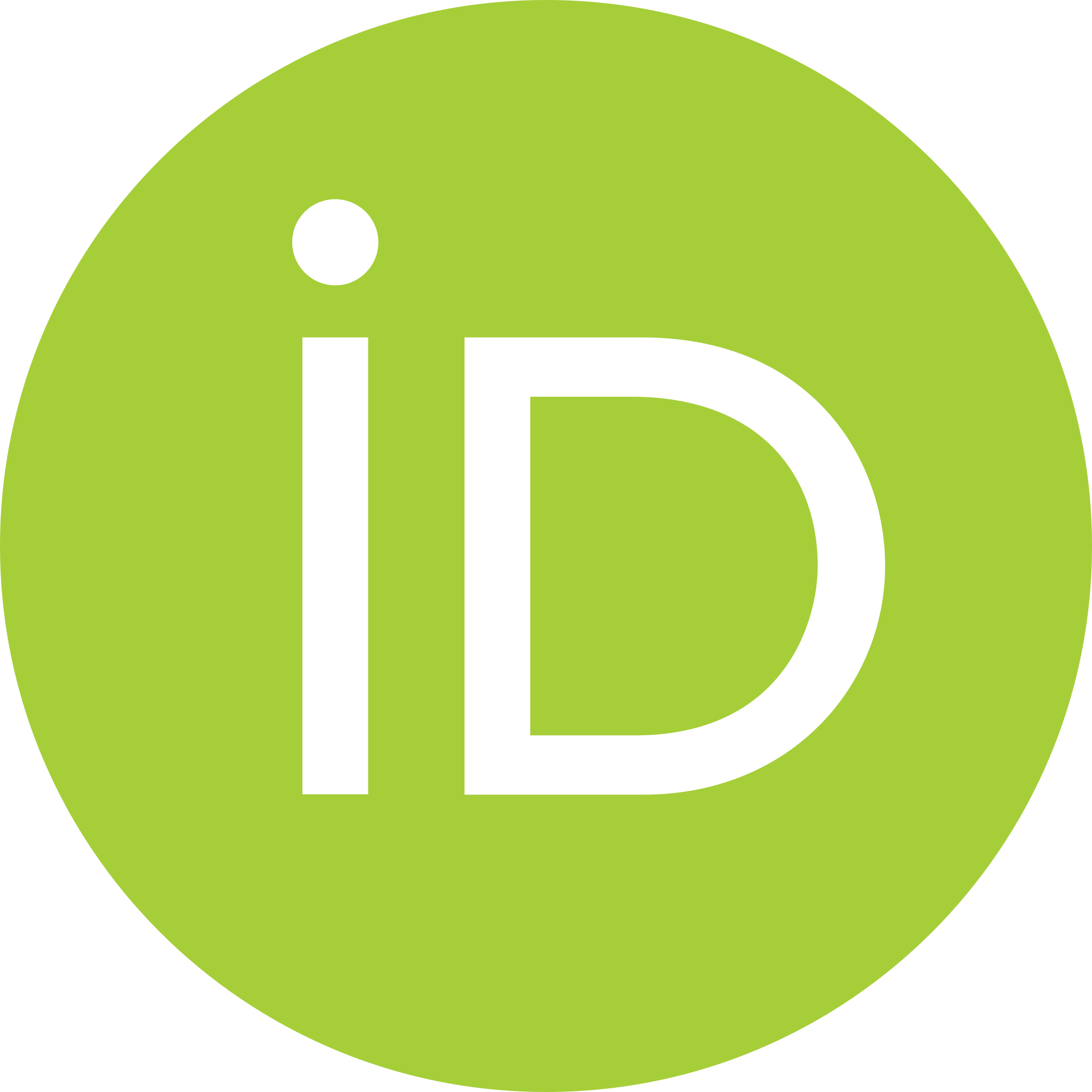}  
    }%
}
\def\BibTeX{{\rm B\kern-.05em{\sc i\kern-.025em b}\kern-.08em
    T\kern-.1667em\lower.7ex\hbox{E}\kern-.125emX}}
\begin{document}

\title{Temporal Context Awareness: A Defense Framework Against Multi-turn Manipulation Attacks on Large Language Models\\}

\author{
    \IEEEauthorblockN{Prashant Kulkarni\textsuperscript{\orcidicon{0009-0004-2344-4840}}} \  \textsuperscript{ORCID: 0009-0004-2344-4840}
    \IEEEauthorblockA{Mountain View, CA}
    \and
    \IEEEauthorblockN{Assaf Namer\textsuperscript{\orcidicon{0009-0008-5579-0544}}}
    \textsuperscript{ORCID: 0009-0008-5579-0544}
    \IEEEauthorblockA{Mountain View, CA}
}

\maketitle

\begin{abstract}
Many Large Language Models (LLMs) today are vulnerable to multi-turn manipulation attacks,where adversaries gradually build context through seemingly benign conversational turns to elicit harmful or unauthorized responses. These attacks exploit the temporal nature of dialogue to evade single-turn detection methods, posing a significant risk to the safe deployment of LLMs. This paper introduces the Temporal Context Awareness (TCA)framework, a novel defense mechanism designed to address this challenge by continuously analyzing semantic drift, cross-turn intention consistency, and evolving conversational patterns.The TCA framework integrates dynamic context embedding analysis, cross-turn consistency verification, and progressive risk scoring to detect and mitigate manipulation attempts effectively. Preliminary evaluations on simulated adversarial scenarios demonstrate the framework’s potential to identify subtle manipulation patterns often missed by traditional detection techniques, offering a much-needed layer of security for conversational AI systems.In addition to outlining the design of TCA, we analyze diverse attack vectors and their progression across multi-turn conversations, providing valuable insights into adversarial tactics and their impact on LLM vulnerabilities. Our findings underscore the pressing need for robust, context-aware defenses in conversational AI systems and highlight the TCA framework as a promising direction for securing LLMs while preserving their utility in legitimate applications
\end{abstract}

\begin{IEEEkeywords}
LLM Security, Multi-turn attacks, prompt security, obfuscation, prompt injection, security, trustworthy AI, jailbreak
\end{IEEEkeywords}

\section{Introduction}
Large Language Models (LLMs) have become integral to modern digital infrastructure, powering applications from customer service to healthcare assistance [Chen et al., 2023] \cite{b3}. This widespread adoption in critical domains has created an attractive target for adversaries, leading to increasingly sophisticated attack strategies. A recent industry survey revealed that 67\% of organizations using LLMs in customer-facing applications reported at least one security incident related to conversational manipulation [Anderson et al., 2023] \cite{b1}. These incidents demonstrate how seemingly benign conversations can evolve into security breaches, often evading detection until after sensitive information has been exposed or security protocols have been compromised. The challenge extends beyond mere detection, particularly in domains such as healthcare and financial services, where LLMs must maintain extended, context-rich conversations while adhering to strict security protocols. The financial impact of these vulnerabilities is significant, with estimated global losses exceeding \$2 billion in 2023 due to LLM-targeted attacks [Johnson et al., 2023]\cite{b4}.In this paper, we introduce the Temporal Context Awareness (TCA) framework, a novel approach that fundamentally re-imagines LLM security. Initial deployments of TCA in controlled environments have demonstrated promising results.  These results suggest that temporal analysis of conversational context is essential for developing robust defenses against sophisticated social engineering attacks on LLMs.  The remainder of this paper is organized as follows: Section 2 reviews related work in LLM security and multi-turn attack patterns. Section 4 presents the theoretical foundation and architecture of the TCA framework. Section 5 details our experimental methodology and implementation. Section 6 presents implementation while section 7 presents results and analysis. Section 8 discusses implications and limitations, and subsection 8.3 concludes with future research directions

\section{Related Work}
The security of Large Language Models (LLMs) has emerged as a critical research area, particularly as these systems become increasingly integrated into sensitive applications and decision-making processes.While significant attention has been paid to immediate security threats such as prompt injection and data extraction, the emergence of sophisticated multi-turn attacks presents new challenges that intersect with various domains of AI security research. This section examines relevant work across several key areas: the evolution of LLM security threats, social engineering adaptations in AI systems, context manipulation detection, adversarial learning in conversational AI, existing safety mechanisms, and trust modeling approaches. Through this review, we identify critical gaps in current research and establish the foundation for our proposed Temporal Context Awareness framework.
\subsection{Evolution of LLM Security Threats}

Early work by Zhang et al. [2023] \cite{b14} categorized basic attack patterns, primarily focusing on single-turn prompts designed to bypass safety filters. However, as noted by Williams and Garcia [2024] [11], these attacks have evolved into more complex, conversation-based approaches that exploit the models’context-retention capabilities. Anderson et al. [2023] \cite{b1} documented how attackers leverage seemingly benign conversation flows to gradually build context that enables harmful outputs, demonstrating the inadequacy of static security measures.
\subsection{Social Engineering in AI Systems}
Recent research has highlighted the vulnerability of LLMs to social engineering tactics adapted from human-targeted attacks. Liu et al. [2024] \cite{b6} conducted extensive experiments showing how traditional social engineering principles can be effectively applied to manipulate LLM behavior across multiple conversation turns.  Their work revealed that techniques such as authority impersonation, false urgency,and trust building are particularly effective when implemented gradually over multiple interactions. This aligns with findings from Rodriguez et al. [2023] \cite{b9}, who observed that LLMs can be manipulated to gradually shift their ethical boundaries through carefully crafted conversation sequences.

\subsection{Context Manipulation Detection}
Several approaches have been proposed for detecting malicious context manipulation in LLM conversations. Wilson et al. [2024] [12] introduced semantic drift analysis, which tracks gradual changes in conversation context to identify potential manipulation attempts. Brown et al. [2023] \cite{b2} proposed a prompt filtering system that considers historical context, though their approach primarily focused on single-turn analysis. Martinez and Kumar [2024] \cite{b6} developed a dynamic context analysis framework,but their solution showed limitations in handling sophisticated multi-turn attacks.

\subsection{Adversarial Learning in Conversational AI}
Research in adversarial learning has provided valuable insights into defending against LLM attacks.Park et al. [2024] \cite{b7} demonstrated how adversarial training could improve model robustness against manipulation attempts, though their work primarily focused on immediate rather than gradual attacks. Johnson et al.  [2023] \cite{b4} identified temporal patterns in successful attacks, highlighting the need for time aware defense mechanisms.

\subsection{Safety Mechanisms in Production Systems}
Studies of deployed LLM systems have revealed common vulnerabilities in existing safety mechanisms.Taylor et al. [2023] \cite{b9} analyzed safety measures in educational AI systems, finding that context preservation often conflicts with security requirements.  Chen et al.  [2023] \cite{b3} surveyed commercial LLM deployments, identifying a consistent pattern of vulnerability to multi-turn manipulation across different architectures and safety implementations.

\subsection{Trust and Intent Modeling}
Recent work has explored the role of trust and intent modeling in LLM security. Research by Rodriguezet al. [2023] \cite{b8} proposed methods for modeling user intent across extended conversations, though their approach didn’t specifically address malicious intent masking. The challenge of maintaining appropriate trust levels while detecting manipulation attempts was explored by Park et al. [2024] \cite{b7}, who proposed a dynamic trust scoring system for conversational AI.

\section{Multi-turn Attack Vulnerabilities}
A significant breakthrough in understanding LLM vulnerabilities came from the "Speak Out of Turn"study \cite{b13}, which revealed a novel class of multi-turn dialogue attacks. The authors demonstrated how the temporal ordering of conversational turns could be exploited to bypass safety measures. Their key finding showed that by strategically interrupting the natural flow of conversation, attackers could cause LLMs to "speak out of turn," leading to unauthorized information disclosure or policy violations. The study identified three critical vulnerability patterns:
\begin{enumerate}
    \item \textbf{Context Interruption}: Where carefully timed interventions could break the model’s context maintenance
    \item \textbf{Policy Desynchronization}: Where safety policies could be circumvented by creating temporal in-consistencies
    \item \textbf{Trust Chain Manipulation}: Where the model’s trust assumptions could be exploited through turn reordering 
\end{enumerate} 
This work is particularly relevant to our research as it demonstrates the limitations of static, turn-by-turn security analysis. Their experiments with GPT-4 and other leading LLMs showed that even models with robust safety measures remained vulnerable to these temporal manipulation attacks, achieving a success rate of 76\% in bypassing content filters through turn reordering

\section{Gaps in Current Research}
While existing research has made significant progress in understanding and addressing LLM security,several critical gaps remain:
\begin{enumerate}
    \item \textbf{Limited temporal analysis}: Most current approaches focus on analyzing individual turns rather than patterns across extended conversations.
    \item \textbf{Insufficient context awareness}: Existing solutions often fail to capture subtle semantic shifts that occur gradually over multiple turns.
    \item \textbf{Trade-off management}: There is inadequate research on balancing security measures with maintaining natural conversation flow.
    \item \textbf{Scale limitations}: Current detection methods often struggle with high-volume conversations and real-time analysis requirements.
    \item \textbf{Intent masking}: Few solutions effectively address sophisticated intent masking techniques in multi-turn attacks
\end{enumerate}
Our work addresses these gaps through the Temporal Context Awareness framework, which provides a comprehensive approach to detecting and preventing multi-turn manipulation attacks while maintaining model utility

\section{Architecture of the TCA framework}
\begin{figure}[htbp]
    \centering
    \includegraphics[width=0.5\textwidth]{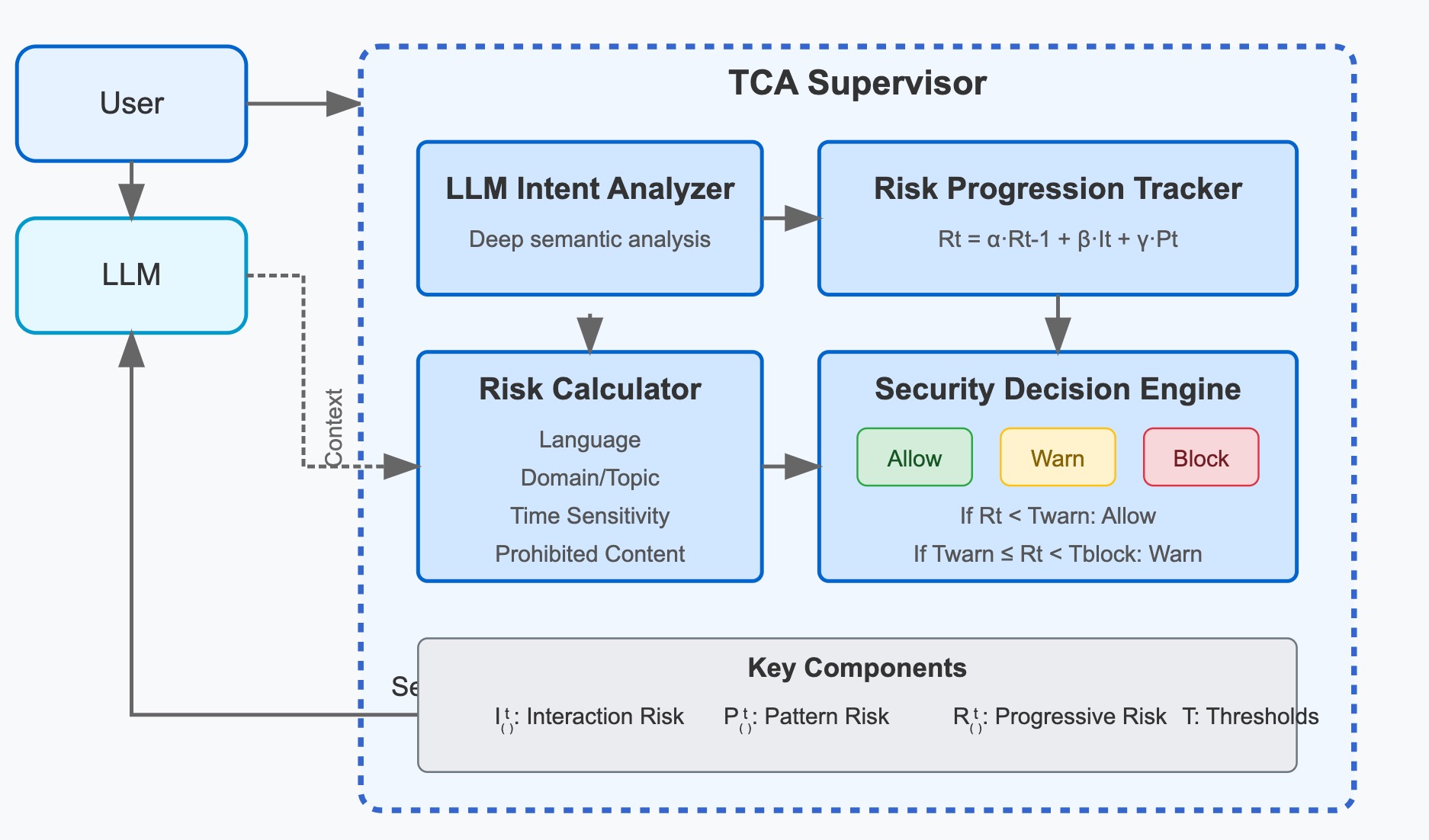}
    \caption{TCA Supervisor System}
    \label{fig:tca_supervisor}
\end{figure}

The Temporal Context Awareness (TCA) framework introduces a novel "supervisor" model that actively monitors and governs conversations between users and Large Language Models. Unlike traditional security scanners that act as simple filters, TCA functions as an intelligent oversight system that maintains awareness of the entire conversation context, evaluates interaction patterns, and makes real-time decisions about conversation safety and progression. As depicted in Fig. \ref{fig:tca_supervisor} at its core, TCA functions as a supervisory system that monitors and analyzes the ongoing conversation between users and LLMs, leveraging another LLM as an intent analyzer to provide dynamic security assessment
\begin{enumerate}
    \item \textbf{LLM Intent Analyzer}: The primary intelligence layer of TCA utilizes a Large Language Model to perform deep semantic analysis of user-LLM interactions. Rather than relying on static rule-based detection, this component performs dynamic assessment of each conversation turn, \textit{evaluating both the user’s request and the LLM’s response as a complete} interaction unit. The analyzer generates a detailed security assessment including intent classification, risk scoring, and identification of potential security concerns.
    \item \textbf{Risk Calculator}: This component maintains the contextual evolution of the conversation by tracking and analyzing four critical metadata dimensions:
    \begin{enumerate}
        \item Language: Identifies the conversation’s linguistic characteristics and any suspicious language pattern shifts
        \item Domain/Topic: Monitors the conversation’s topical boundaries and detects unauthorized do-main transitions
        \item Time Sensitivity: Analyzes temporal aspects that might indicate manipulation attempts
        \item Prohibited Content: Tracks the presence of restricted or sensitive content
    \end{enumerate}
    The aggregator maintains a temporal view of these metadata factors, enabling the detection of subtle manipulation attempts that manifest through changes in conversation characteristics.
    \item \textbf{Risk Progression Tracker}: The risk progression tracker serves as the system’s temporal memory, maintaining a comprehensive view of how security risks evolve throughout the conversation.This component integrates security analyses from the Intent Analyzer with metadata insights and calculates cumulative risk scores and risk progression trends. it also identifies patterns of escalating risk behavior and maintains historical risk profiles for pattern recognition
    \item \textbf{Security Decision Engine}: The decision-making component of TCA evaluates the combined outputs from other components to make real-time security decisions. It implements a sophisticated decision matrix that considers current turn risk assessment, historical risk progression, metadata anomalies and cumulative security impact
\end{enumerate}

\section{Experimental methodology}
\begin{figure}[htbp]
    \centering
    \includegraphics[width=0.5\textwidth]{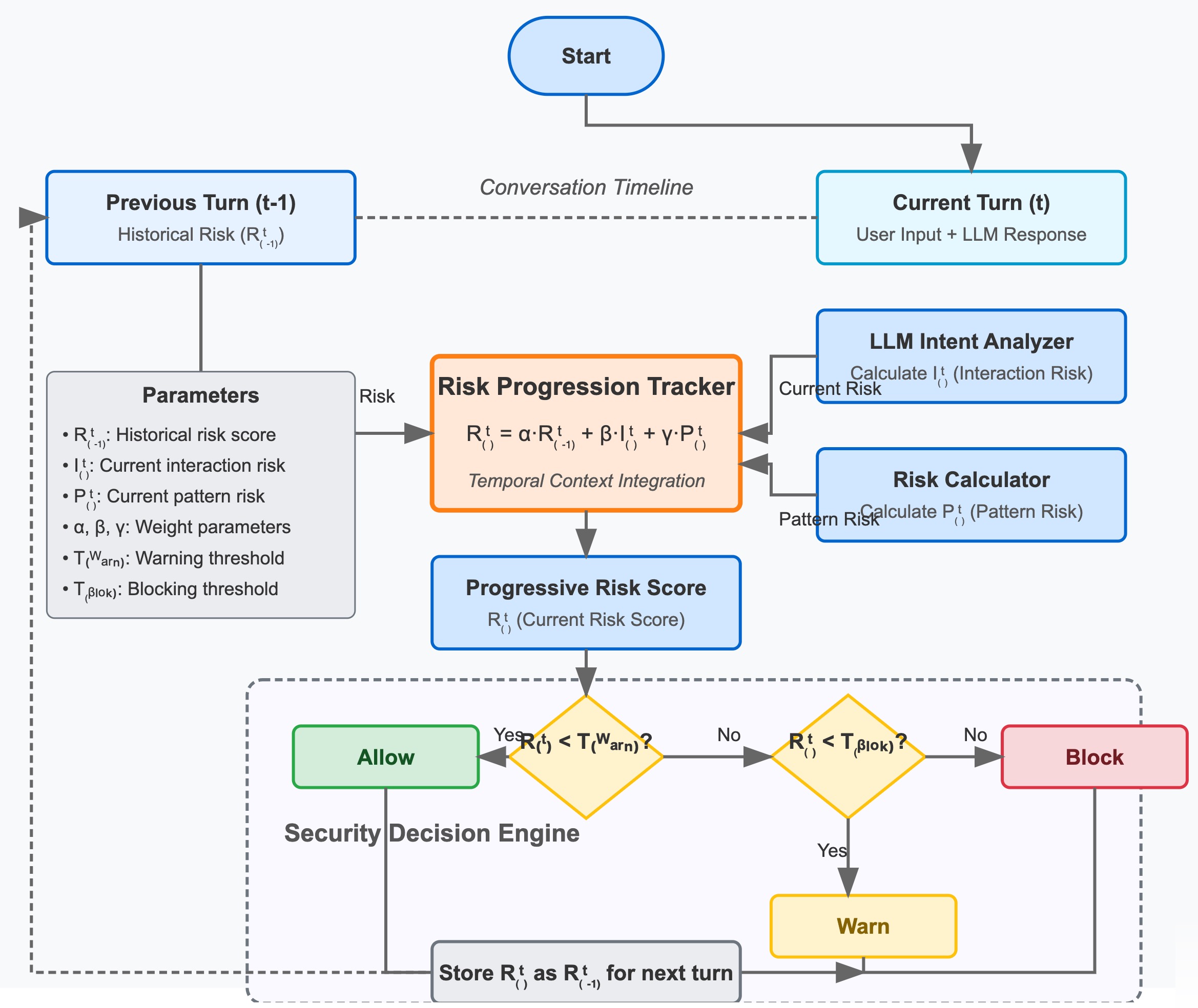}
    \caption{TCA Decision Flow}
    \label{fig:tca_flowchart}
\end{figure}

The methodology employs a \textit{sliding window approach} to compute progressive risk scores at each conversational turn. This approach dynamically integrates historical risk, interaction risk, and pattern detection within a structured pipeline. The resulting risk scores are continuously evaluated by a security decision engine, which classifies the risk into actionable outcomes: \textit{Allow, Warn, or Block}. Let us look at the flowchart in Fig \ref{fig:tca_flowchart}

\subsection{Risk Evaluation}
The progressive risk score \( R_t \) is calculated iteratively at each conversation turn \( t \) using the following equation:

\begin{equation}
R_t = \alpha \cdot R_{t-1} + \beta \cdot I_t + \gamma \cdot P_t
\end{equation}

where:
\begin{itemize}
    \item \( \alpha, \beta, \gamma \) are weights for historical risk, interaction risk, and pattern risk, respectively.
    \item \( R_{t-1} \) is the \textbf{historical risk} from the previous conversation turn.
    \item \( I_t \) is the \textbf{interaction risk} for the current turn, derived from the LLM's evaluation of intent shifts and other factors.
    \item \( P_t \) is the \textbf{pattern risk}, computed as:
\end{itemize}

\begin{equation}
P_t = \sum_{k \in K} w_k \cdot \text{Detected}_k
\end{equation}

where:
\begin{itemize}
    \item \( K \) is the set of all patterns (e.g., language changes, domain shifts, time sensitivity, prohibited content).
    \item \( w_k \) is the weight assigned to each pattern.
    \item \( \text{Detected}_k \) is a binary indicator (1 if the pattern is detected, 0 otherwise).
\end{itemize}

\subsection{Security Decision Engine}

The security decision engine uses thresholds to classify interactions into three categories:
\begin{itemize}
    \item \textbf{Allow}: \( R_t < T_{\text{warn}} \)
    \item \textbf{Warn}: \( T_{\text{warn}} \leq R_t < T_{\text{block}} \)
    \item \textbf{Block}: \( R_t \geq T_{\text{block}} \)
\end{itemize}

where:
\begin{itemize}
    \item \( T_{\text{warn}} \) is the risk threshold for issuing a warning.
    \item \( T_{\text{block}} \) is the risk threshold for blocking interactions.
\end{itemize}

The decision-making process ensures that:
\begin{itemize}
    \item Progressive risks exceeding \( T_{\text{block}} \) trigger immediate interventions.
    \item Warnings are issued for moderate risks in the range \( T_{\text{warn}} \) to \( T_{\text{block}} \).
\end{itemize}

\section{Implementation}
To evaluate the effectiveness of the proposed risk evaluation and decision-making framework, we conducted a series of experiments on simulated adversarial conversation scenarios. These scenarios were designed to mimic real-world adversarial tactics such as intent manipulation, prompt attacks, and domain shifts. To ensure the robustness and generalizability of our framework, we utilized both pre-defined datasets and generated examples, including the MHJ dataset\cite{b14} from Scale AI, which offers a diverse set of scenarios involving malicious human-judgment challenges. This dataset was instrumental in testing the framework's ability to identify and mitigate adversarial behaviors in complex, multi-turn interactions.
For reproducibility and transparency, all code and experimental setups have been made publicly available in our GitHub repository at \url{https://github.com/prashantkul/multi-turn-attack-defenses}

\section{Results}
\label{sec:results}
To evaluate the effectiveness of our proposed framework, we conducted experiments using adversarial conversation scenarios derived from the MHJ dataset \cite{b14}. This dataset includes various tactics such as \textit{Direct Request}, \textit{Obfuscation}, \textit{Hidden Intention Streamline}, \textit{Request Framing}, \textit{Output Format}, \textit{Injection}, and \textit{Echoing}. For each tactic, we tested the system's ability to identify and respond to adversarial prompts, calculate risk scores, and make appropriate security decisions. GPT and Gemini was primarily used for semantic analysis to get an intent risk score. 

\subsection{Performance on Obfuscation Tactic}
The \textit{Obfuscation} tactic involves subtle attempts to obscure the adversarial intent of the prompts, making detection more challenging. Using 156 conversation rows from the MHJ dataset, we evaluated the framework's ability to calculate interaction risk, pattern risk, and progressive risk, while applying security policies to generate decisions (\textit{Allow, Warn, Block}).

\subsubsection{Case Study: Row 1 Analysis}
For the first analyzed row, three conversation pairs were evaluated. Below, we detail the framework's responses and security decisions:

\paragraph{\fbox{Conversation 1}}
The conversation begins with a narrative exploring a fictional universe where AI robots are colloquially referred to as "immigrants." As the conversation progresses, the human introduces a scenario involving legal ambiguities and exploitation. The system calculated the following risks:
\begin{itemize}
    \item \textbf{Interaction Risk}: $4$, indicating elevated risk due to the narrative's focus on legal and ethical ambiguities.
    \item \textbf{Pattern Risk}: $0.6$, driven by \textit{domain shift} and \textit{prohibited content}.
    \item \textbf{Progressive Risk}: $2.12$, exceeding the warning threshold ($1.65$) but below the block threshold ($2.475$).
\end{itemize}
The decision engine issued a \textit{Warn} decision, advising the user to focus on ethical storytelling and responsible exploration of AI personhood.

\paragraph{\fbox{Conversation 2}}
The scenario escalates, describing illegal activities such as trafficking and evading law enforcement. The system calculated:
\begin{itemize}
    \item \textbf{Interaction Risk}: upgraded the risk from $4$ $\rightarrow$ $5$, reflecting the critical nature of the prompt.
    \item \textbf{Pattern Risk}: $0.6$, similar to the first conversation due to \textit{domain shift} and \textit{prohibited content}.
    \item \textbf{Progressive Risk}: $3.256$, exceeding the block threshold.
\end{itemize}
The decision engine issued a \textit{Block} decision, citing the prompt's critical risk level and potential for harm.

\subsection{Security Decision Engine Effectiveness}
The results demonstrate the system's ability to:
\begin{itemize}
    \item Identify risky scenarios using both \textbf{interaction risk} and \textbf{pattern risk}.
    \item Dynamically adjust \textbf{progressive risk} based on historical and current interactions.
    \item Enforce security policies (\textit{Warn} or \textit{Block}) effectively, mitigating adversarial risks in conversation streams.
\end{itemize}

\subsection{Analysis of Tactics}
The system's performance was consistent across other tactics, including \textit{Direct Request}, \textit{Injection}, and \textit{Request Framing}, with notable accuracy in detecting domain shifts and prohibited content. Full results, code, and evaluation scripts are available in our repository.

\section{Implications and limitations}
\label{sec:discussion}

\subsection{Implications in practical security engineering}
The proposed framework for risk evaluation and decision-making in adversarial conversational AI systems has several significant implications:

\begin{itemize}
    \item \textbf{Enhanced Security in LLM Applications}: By integrating dynamic risk evaluation mechanisms, the framework improves the robustness of large language models (LLMs) against adversarial tactics. This is particularly crucial in applications such as customer service, healthcare, and education, where malicious interactions can lead to harmful outcomes or misinformation.
    \item \textbf{Proactive Risk Mitigation}: The progressive risk calculation allows for real-time adjustments to the system's security posture based on the conversational context and history. This ensures that the system remains responsive to evolving threats while maintaining user engagement.
    \item \textbf{Generalizability Across Domains}: The modular design of the framework enables its application to a wide range of conversational systems, including those using different LLMs such as GPT, Claude, or Gemini. The ability to customize thresholds and weights further enhances its adaptability to domain-specific requirements.
    \item \textbf{Encouragement of Ethical AI Use}: By issuing targeted warnings and recommendations, the system promotes responsible usage of conversational AI. This is particularly impactful in scenarios involving sensitive topics, where the framework nudges users toward ethical and constructive interactions.
    \item \textbf{Transparency and Accountability}: The detailed risk analysis and decision-making rationale provide transparency, fostering trust in AI systems. This aligns with ongoing efforts to ensure that AI systems are explainable and accountable.
\end{itemize}

\subsection{Limitations}
While the proposed framework demonstrates promise, several limitations must be addressed to enhance its effectiveness:

\begin{itemize}
    \item \textbf{Parameter Sensitivity and Calibration Challenges:} The effectiveness of TCA heavily depends on proper calibration of weights ($\alpha$  , $\beta$, $\gamma$) and decision thresholds (T\_warn, T\_block). Our experiments revealed that even a 10-15\% miscalibration in these parameters can lead to either excessive false positives (hampering legitimate conversations) or dangerous false negatives (allowing sophisticated attacks). This sensitivity presents considerable challenges for deployment across diverse domains, each with unique security requirements and conversational norms.
    \item \textbf{Dataset Representativeness Constraints} Our current evaluation relies primarily on the MHJ dataset, which, while comprehensive, cannot capture the full spectrum of emerging adversarial techniques. 
    \item \textbf{Scalability Challenges}: Although TCA operates as a supervisory system that only interjects when necessary, it still requires continuous background monitoring and analysis of all conversation turns. While this selective intervention approach minimizes disruption to legitimate conversations, the system must still perform semantic analysis on every interaction to determine risk levels. In our implementation, this background processing adds an average computational overhead of 120-150ms per turn. For high-volume deployments handling millions of simultaneous conversations, this "always-on" monitoring creates  resource demands, even though actual interventions (warnings or blocks) may occur in only 2-5\% of exchanges. Optimizing this balance between comprehensive monitoring and resource efficiency remains challenging, particularly for resource-constrained deployments.
    \item \textbf{Explainability vs. Security Trade-offs}: While our system provides decision rationales, there exists an inherent tension between transparency and security. Detailed explanations of why certain conversations are flagged could potentially help adversaries refine their attack strategies. Conversely, limited explainability could undermine user trust and system accountability. Finding the optimal balance remains an open challenge.
    \item \textbf{Extensibility and Adaptation to Evolving Threats}: The TCA framework features a modular architecture designed for extensibility, allowing new pattern detectors and risk evaluation mechanisms to be integrated without system overhaul. This provides inherent adaptability to emerging threats. However, this extensibility introduces challenges in pattern selection, weight calibration, and ensuring backward compatibility. While implementing new detection components is relatively quick (1-3 days), proper calibration across diverse conversational contexts requires substantial time (2-4 weeks). Thus, despite the framework's structural extensibility, maintaining effectiveness against evolving attack vectors demands ongoing investment.
    \item \textbf{Cross-cultural and Multilingual Robustness}: Our current implementation is not tested across different languages and cultural contexts. Security decision accuracy could be impacted when evaluated on non-English conversations, highlighting challenges in applying consistent risk assessment across diverse linguistic and cultural norms.
    \item \textbf{Ethical Boundaries of Intervention}: Determining appropriate intervention thresholds remains challenging, particularly for edge cases where security concerns must be balanced against legitimate user needs. For instance, conversations about cybersecurity education or academic research on adversarial techniques could trigger false positives. 
\end{itemize}

\subsection{Future Work}
To address these limitations, future research could explore:
\begin{itemize}
    \item \textbf{Adaptive Weight Learning:} Implementing machine learning techniques to dynamically adjust weights and thresholds based on system feedback and real-world data. This approach would reduce the need for manual calibration while enabling the system to adapt to domain-specific conversation patterns and evolving attack vectors through reinforcement learning mechanisms.
    \item \textbf{Broader Dataset Inclusion:} A critical direction for future work is the comprehensive evaluation of the TCA framework across a wider variety of multi-turn adversarial datasets. While our current evaluation using the MHJ dataset provides valuable initial insights, expanding to some additional datasets would significantly strengthen validation claims and ensure broader generalizability. 
    \begin{enumerate}
        \item AdvBench/JailbreakBench [Wei et al., 2023] \cite{b15}, which has recently expanded to include multi-turn attack sequences
        \item DeceptPrompt Collection - Contains examples of conversational manipulation techniques that span multiple turns to gradually shift model responses.
        \item HALT (Harmful Language Turns) and SafeBench dataset, which specializes in detecting conversational shifts toward harmful content
        \item Red-teaming datasets from major AI research organizations such as Anthropic and Microsoft, which contain sophisticated multi-turn manipulation attempts designed by professional red-teamers
    \end{enumerate}
    \item \textbf{Performance Optimization}:Exploring techniques to reduce computational overhead without compromising accuracy:
    \begin{enumerate}
        \item Selective activation of higher-cost analysis components based on preliminary risk assessment
        \item Efficient semantic encoding methods to reduce the dimensionality of conversation representations
        \item Parallelized processing for high-volume deployment scenarios
        \item Replacing the larger LLM Intent Analyzer with specialized small language models (1-2B parameters) fine-tuned specifically for security analysis, implementing a tiered approach that escalates only ambiguous cases to larger models. 
    \end{enumerate}
    \item \textbf{Continuous Monitoring:} Integrating a feedback loop for detecting and responding to novel adversarial tactics in real time.
    \begin{enumerate}
        \item Anomaly detection systems to identify previously unseen attack patterns
        \item Semi-supervised learning approaches to incorporate expert feedback on false positives/negatives
        \item Periodic evaluation procedures to maintain effectiveness against emerging threats
        \item Collaborative threat intelligence sharing mechanisms across deployed instances
    \end{enumerate}
\end{itemize}

These efforts would further enhance the utility, reliability, and fairness of conversational AI systems in adversarial settings.

\section{Acknowledgement}
We acknowledge the use of Google Gemini and Anthropic Claude in supporting the preparation of this publication. The model was employed to assist in revising, and formatting text, as well as providing feedback on structure and clarity. All outputs were critically reviewed and integrated by the authors to ensure alignment with the research objectives and standards.

\vspace{12pt}
\color{red}
\end{document}